\documentclass[12pt]{article}
\usepackage{geometry}
\usepackage{graphicx}
\usepackage{latexsym}
\usepackage{amsmath}
\usepackage{amsthm}
\usepackage{amssymb}
\usepackage[english]{babel}
\usepackage{algorithm}
\usepackage{algpseudocode}

\begin{document}
  \begin{flushleft}
  {\footnotesize \bf Gakuto International Series}\\
  {\footnotesize \bf Mathematical Sciences and Applications Vol.XX (2011)}\\
  {\footnotesize \it International Symposium on Computational Science 2011}\\
  {\footnotesize pp.XXXXX} 
  \end{flushleft}
  \vspace{1.5cm}
                   \begin{center}
   {\large \bf 
         Stochastic moving particle semi-implicit for inviscid fluid wave simulation}\\[1cm]
                   \end{center}

                   \begin{center}
                  {\sc Christian Fredy Naa}\\
			Institute of Science and Engineering, School of Mathematics and Physics, Kanazawa University\\
			Faculty of Mathematics and Natural Science, Institut Teknologi Bandung\\
\vspace{0.2cm}

 		{\sc Seiro Omata}\\
			Institute of Science and Engineering, School of Mathematics and Physics, Kanazawa University\\
\vspace{0.2cm}

		{\sc Masaki Kazama}\\
			Fujitsu Limited\\
                    \end{center}
\vspace{2cm}
         
\noindent
{\bf Abstract.} The present paper introduces stochastic velocity as improvement for moving particle semi-implicit (MPS) method. This improvement is to overcome energy loss caused by numerical dissipation in the basic MPS that brings about rapid decay of waves.  Stochastic velocity is added in the explicit step of the basic MPS method. MPS with stochastic improvement is compared with the basic method in the case of linear water waves, in particular dam break problem and standing wave in a rectangular tank. Surface detection and curve fitting are used to analyze the parameters of wave on the standing wave case. The surface detection and curved fitting was efficient to determine parameters of the wave and it was found that the stochastic improvement made the waves survived longer than in the basis method.

        \vspace{4cm}
         
\vfill
\noindent
--------------------------------------------------------------------
----------
\\
{\footnotesize Received xxxxxxxxxx, 2011.\\
This work is supported by xxxxxxxxxxxxxxxxxxxx.\\
AMS Subject Classification xxxxxxxxxx, xxxxxxxxxx} 

\newpage

\noindent
{\section{Introduction}
Moving particle semi-implicit (MPS) \cite{[kos1]} method is a particle method for simulating incompressible fluids. MPS method was used for example to analyze breaking waves \cite{[kos3]}, droplet breakup behavior \cite{[nomura]} and to predict wave impact pressure \cite{[khayyer]}. Since MPS method is based on Lagrangian system, computational grids are  not necessary. Governing equations are discretized based on particle interaction models representing density, gradient, Laplace operator and free surface. 

However, MPS has a weakness in the energy conservation. The waves calculated by the basic MPS decay rapidly since the mechanical energy is not fully conserved. This lack of conservation of energy is caused by numerical dissipation. A work has been introduced to recover this weakness: Suzuki \cite{[suzuki]} introduced Hamiltonian moving particle semi-implicit (HMPS). Even then, the mechanical energy was not fully conserved, but the HMPS was able to make the waves survive longer.

The objective of this paper is to overcome the loss of energy in the basic MPS using stochastic modification of velocity. The purpose of this stochastic concept is to add extra kinetic energy to the particles so that the kinetic energy gained by the stochastic velocity recovers the loss of energy caused by numerical dissipation.

The continuity equation and Euler's equation are used as governing equation. Stochastic and basic MPS were compared in the cases of dam break problem and standing wave in a rectangular tank.

Surface detection algorithm is used to determine amplitude in the standing wave case. This kind of algorithm is used because it is hard to judge the important parameters of the wave only from the distribution of particles. 

The paper is organized in the following way. A brief explanation of standard MPS method is presented in second section. The stochastic improvement of velocity is introduced in third section. The fourth section describes the surface detection algorithm. Finally, the improved MPS method compared with the standard MPS method in the case of dam break problem and standing wave in a rectangular tank.

\section{Standard MPS method}

In this section, the MPS method is briefly explained based on description which provided by Koshizuka \cite{[kos1]}. In the MPS method, the fluid is modeled using interaction of particles according to equations of motion. Governing equations for inviscid fluid motion are continuity equation and Euler's equation:
	\begin{equation}
		\frac{1}{\rho}\frac {D \rho}{D t} + \nabla \cdot {\bf{u}} = 0
		\label{eq:continuity}	
	\end{equation}
	\begin{equation}
		\frac{D\bf{{u}}}{Dt} = -\frac{1}{\rho} \nabla P +  {\bf{g}}, 
		\label{eq:euler}
	\end{equation}
where ${\bf{u}}$ denotes particle velocity vector, $t$ denotes time, $\rho$ denotes density, $P$ denotes pressure and ${\bf{g}}$ denotes gravity acceleration vector. 

It should be noted that Eq. (\ref{eq:continuity}) is written in the form of a compressible flow. In the MPS method, incompressibility is enforced by the way setting $\frac{D\rho}{Dt}=0$ at each particle at each calculation time step. According to \cite{[khayyer]} the left hand side of Eq. (\ref{eq:euler}) denotes the material derivative $\frac{D\bf{{u}}}{Dt}$ involving the advection term. In the particle methods, including the MPS method, the advection term is automatically calculated through the tracking of particle motion; hence, the numerical diffusion arising from the successive interpolation of the advection function in Eulerian grid based methods is controlled without the need for a sophisticated algorithm. 

The main idea of MPS method is to divide Eq. (\ref{eq:euler}) into two parts to calculate the change of velocity, as follows
	\begin{equation}
	\label{eq:mps1}
	\left(\frac{d{\bf{u}}}{{dt}}\right)^{\text{explicit}} =  {\bf{g}},
	\end{equation}
	\begin{equation}
	\label{eq:mps2}
	\left(\frac{d{\bf{u}}}{dt}\right)^{\text{implicit}} = -\frac{1}{\rho}\nabla P.
	\end{equation}

Particle interaction is described in terms of weight function. Weight function $w(r)$ in MPS method is defined as \cite{[kos1]}
	\begin{equation}
	w(r) = 
		\begin{cases}
		\frac{r_e}{r}-1  & \text{if } r < r_e
		\\0 & \text{if } r \ge r_e.
		\end{cases}
	\label{eq:kernel}
	\end{equation}
Here, $r$ will have the meaning of distance between particles and $r_e$ is the \emph{cut off} distance. In this paper, $r_e$ equals to $2.1 l_0$, where $l_0$ is the initial distance between particles.

The particle number density for particle $i$ ($n_i$) is calculated by 
	\begin{equation}
	n_i = \sum_{j\neq i}^N w(r),
	\label{eq:pnd}
	\end{equation}	
where $N$ denotes the total number of particles. 

The Laplace operator of a scalar quantity $\phi$ for particle $i$ is evaluated using 
	\begin{equation}
	\left<\nabla^2 \phi\right>_i = \frac{2d}{\lambda n_0}\sum_{j\neq i}^N (\phi_j - \phi_i) w(|{\bf{r}}_j- {\bf{r}}_i|).
	\label{[eq:laplacian]}
	\end{equation}
Here, ${\bf{r_i}}$ is position vector of particle $i$ and $d$ is the space dimension. In this paper, the constant $n_0$ is defined as the maximum value of particle number density
	\begin{equation} 
	n_0 = \max{n_i}.
	\label{eq:nzero}
	\end{equation}
The parameter $\lambda$ is defined as \cite{[khayyer]}
	\begin{equation}
	\lambda = \frac{\sum_{i \neq j}^N |{\bf{r}}_j- {\bf{r}}_i|^2 w(|{\bf{r}}_j- {\bf{r}}_i|)} {\sum_{i \neq j}^N w(|{\bf{r}}_j- {\bf{r}}_i|)}.
	\label{eq:lambda}
	\end{equation}
The gradient of pressure $\nabla P$ for particle $i$ is defined as
	\begin{equation}
	\left<\nabla P\right>_i = \frac{d}{n_0}\sum_{j \neq i}^N \frac{P_j - \hat{P}_i}{|{\bf{r}}_j- {\bf{r}}_i|^2}w(|{\bf{r}}_j- {\bf{r}}_i|)({\bf{r}}_j - {\bf{r}}_i),
	\label{eq:pgradmod}
	\end{equation}
where $\hat{P}_i$ denotes the minimum pressure among particles within certain cut-off distance.

For modeling the incompressibility, the number of densities $n^*$ that are calculated at the end of explicit step deviate from the constant number of density $n_0$; hence, a second corrective process is required to adjust the number of densities to initial values prior to the time step. In the implicit step, the intermediate particle velocities are updated implicitply through solving the Poisson Pressure Equation (PPE) derived as \cite{[kos1]}
	\begin{equation}
	\left<\nabla^2 P_{k+1}\right>_i = \frac{\rho}{\Delta t^2} \frac{n_0-(n_k^*)_i}{n_0},
	\label{eq:comp5}
	\end{equation}
where $\Delta t$ denotes calculation time step and $k$ denotes the step of calculation.

To set the Dirichlet boundary condition for the Poisson's equation of pressure, particles satisfying \cite{[nomura]}
	\begin{equation}
	n_i < \beta n_0,
	\label{eq:freesurface}
	\end{equation}
are judged as surface particles and their pressure is fixed to zero or to atmospheric pressure value. Here, $\beta$ is a value between $0.8$ to $0.98$.

The time step is important for numerical stability. According to \cite{[szhang]} the time step should follow the CFL condition
	\begin{equation}
	\Delta t \leq 0.2 \frac{l_0}{u_{\text{max}}},
	\label{eq:courant}
	\end{equation}
here, $u_{\text{max}}$ is the maximum velocity among the particles.

\section{Modification of MPS method}

As described before that the standard MPS method has a weakness in the energy conservation. The mechanical energy is not fully conserved caused by numerical dissipation. To overcome this problem, the stochastic improvement of velocity is introduced. The stochastic velocity is added in standard MPS method. 

To add such stochastic velocity, after the particle's velocity in the explicit step (Eq. (\ref{eq:mps1})) is calculated, the stochastic velocity ${\bf{u}}^{\text{stoc}}$ is added. The direction of stochastic velocity ${\bf{u}}^{\text{stoc}}$ should be determined so that it does not reduce the particle velocity in the particular time step. To generate such direction, first the unit vector of particle's velocity {\bf{u}} is calculated by
	\begin{equation}
	\begin{cases}
	\hat{u}_x = \frac{u_x}{|{\bf{u}}|} \\
	\hat{u}_y = \frac{u_y}{|{\bf{u}}|},
	\end{cases}
	\label{eq:uvecstoc}
	\end{equation}
where $|{\bf{u}}|$ is the length of the vector. Next, random angle $\theta$ is generated from the interval ($-\pi / 2$,$\pi/2$) by choosing a random number $r$ from a uniform distribution $U(0,1)$:
	\begin{equation}
	\theta = r\left(\pi/2-(-\pi/2)\right)-\pi/2.
	\label{eq:trand}
	\end{equation}

Then, rotating the vector $\hat{u}_x$ and $\hat{u}_y$ by the angle $\theta$ counter-clockwise about the origin, the unit vector of the stochastic velocity is determined by
	\begin{equation}
	\left(\begin{array}{cc}
	u_x^{\text{stoc}} \\ 
	u_y^{\text{stoc}}
	\end{array}\right) = 
	\left(\begin{array}{cc}
	\cos\theta & -\sin\theta \\ 
	\sin \theta & \cos \theta
	\end{array}\right)
	\left(\begin{array}{cc}
	{\hat{u}_x} \\ 
	{\hat{u}_y}
	\end{array}\right)
	\label{eq:trans}
	\end{equation} 	
Finally, the unit vector determined by Eq. (\ref{eq:trans}) is multiplied by the magnitude of the stochastic velocity $m$ to get the final vector form 
	\begin{equation}
	{\bf{u}}^{\text{res}} = {\bf{u}} + m {\bf{u}}^{\text{stoc}}.
	\label{eq:mrand}
	\end{equation}
Here, {\bf{u}} is the velocity calculated by basic MPS method (after the explicit step). ${\bf{u}}^{\text{res}}$ is the resultant velocity due to addition ${\bf{u}}$ and ${\bf{u}}^{stoc}$ This process is illustrated in Fig. \ref{fig:stoc}.

	\begin{figure}[hbt]
	\centering
	\includegraphics[scale = 0.75]{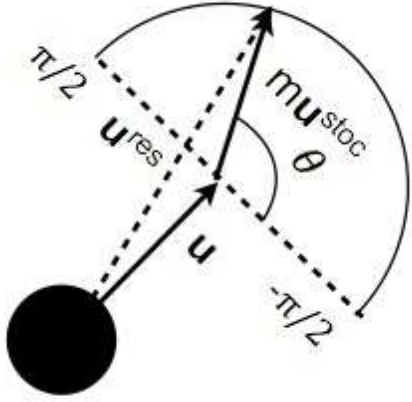}
	\caption{Stochastic velocity}
	\label{fig:stoc}
	\end{figure}

The magnitude of the stochastic velocity $m$ is also given by choosing a random number $r$ from a uniform distribution $U(0,1)$:
	\begin{equation}
	m = r\left(\text{max}-\text{min}\right)+\text{min},
	\label{eq:mrand}
	\end{equation}
where $\text{min}$ is the minimum range of the magnitude which is defined to be $0$, $\text{max}$ is the maximum range which defined as
	\begin{equation}
	\text{max} = \alpha u_{\text{loss}}.
	\label{eq:mag}
	\end{equation}
Here, $\alpha$ is a positive constant which determines the strength of the stochastic velocity, while the term $u_{\text{loss}}$ is defined as the amount of velocity that is lost due to energy difference on subsequent time steps. 

The value of $u_{\text{loss}}$ is determined based on the idea that the energy loss on two neighboring time steps corresponds to the loss of kinetic energy $\Delta EK$
	\begin{equation}
	\Delta EK = E_{\text{total}}^{k-1} - E_{\text{total}}^{k},
	\label{eq:maxm}
	\end{equation}
where $E_{\text{total}}$ denotes total energy and $k$ denotes step number. Since kinetic energy per unit mass is given by $EK = \frac{{\bf{|u|}}^2}{2}$, then 
	\begin{equation}
	u_{\text{loss}} = \sqrt{2\Delta EK}.
	\label{eq:vloss}
	\end{equation}	
The addition of stochastic velocity follows several constraints:
\begin{enumerate}
\item{From experience of several simulations, the constant $\alpha$ is usually much less than $0.1$. If its more than $0.1$, the particles will get extra high velocity that leads to excessive energy making the particles jump off.}
\item{It may happen that $\Delta EK$ in Eq. (\ref{eq:maxm}) equals to zero or even to a negative number (which means the energy is larger than in the previous step). In this case, the explicit step will be calculated without stochastic velocity.}
\end{enumerate}

The complete algorithm of stochastic MPS in some step $k$ is described below: 

\begin{algorithm}
           \caption{Stochastic MPS method algorithm}\label{stocalgo}
		\begin{algorithmic}[1]
			\Require Total energy $E^{k-1}_{total}$ 
			\State Calculate total energy $E^{k}_{\text{total}}$.
			\State Determine direction and magnitude ${\bf{u}}^{\text{stoc}}$
			\If {$E^{j-1}_{\text{total}} \le E^{k}_{\text{total}}$}
				\State ${\bf{u}}^{\text{res}} = {\bf{u}} + m {\bf{u}}^{\text{stoc}}$
			\Else 
				\State ${\bf{u}}^{\text{res}} = {\bf{u}}$
			\EndIf
			\State $E^{k-1}_{\text{total}} = E^{k}_{\text{total}}$.
			\State check termination.
		\end{algorithmic}
\end{algorithm}

This paper concerns the addition of stochastic velocity in 2D case; similar idea can be adopted for 3D case. Since this stochastic concept is a new improvement in MPS method, it is not yet verified by any other research. 

\section{Surface Detection} 

In order to compare the result from the simulations of basic MPS method and the  stochastic improvement, a surface detection algorithm is needed. It is hard to judge the important parameters of the wave only from the distribution of particles.

The algorithm begins with finding the upper left and right particles. Then the algorithm starts to trace the surface particles from the left to the right. From the left particle, the vertical reference axis is used to detect the second surface particle. 
	\begin{figure}[hbt]
	\centering
	\includegraphics[scale = 0.75]{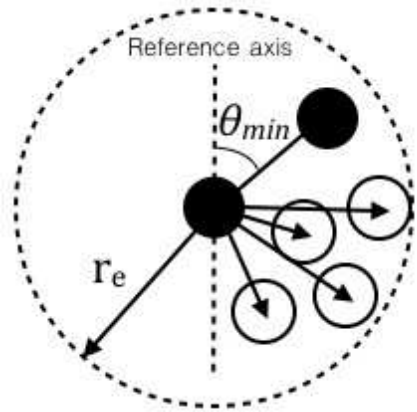}
	\caption{Detection by using vertical imaginary axis. The solid circles denote the surface solution.}
	\label{fig:fdetect}
	\end{figure}

The mechanism to detect the second surface particles is shown in Fig. \ref{fig:fdetect}. The radius of search area is bounded by constant $r_e$ which is the same as the cut-off radius in the above described theory of MPS. Angle from the vertical reference axis to each particle within the search radius is measured. Once the angles are measured, they are converted into quadrant value. 
The second surface particle is then determined as the particle that has the minimum angle with the vertical reference axis. After the second surface particle is determined, this surface particle is called the \emph{current point} and it is used to get the third surface particle. 
	\begin{figure}[!hbt]
	\centering
	\includegraphics[scale = 0.45]{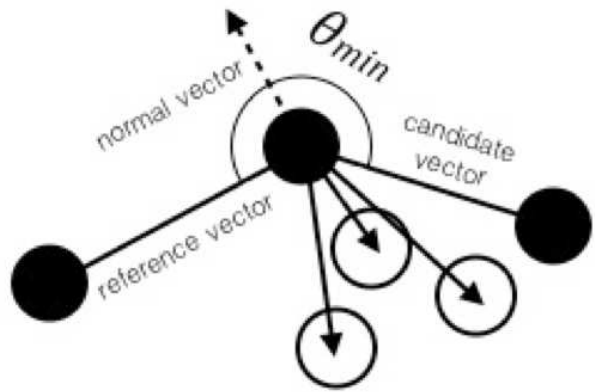}
	\caption{Detection using dot product relation. The solid circles denote the surface solution.}
	\label{fig:dotrelation}
	\end{figure}

The third surface particle and so on are determined using different algorithm. This algorithm is shown in Fig. \ref{fig:dotrelation}, the surface particle detected before the current point is used (in this case: the third surface particle is determined using the first and the second particles).

First, the reference vector that connects the current point and the previous surface particle is measured. Then the normal vector perpendicular to the reference vector which direct out from the particle distribution is determined using geometry transformation matrix. Candidate vectors from current point to all the particles inside the search area are also measured.

Using two dot products relation (the product of between the reference vector and the candidate vectors, and the product of normal vector and the candidate vectors), the angles $\theta$ from the reference vector to the candidate vectors as shown in Fig. \ref{fig:dotrelation} can be determined. The next surface particle is the one corresponding to the smallest angle $\theta$.
	\begin{figure}[hbt]
	\centering
	\includegraphics[scale = 0.6]{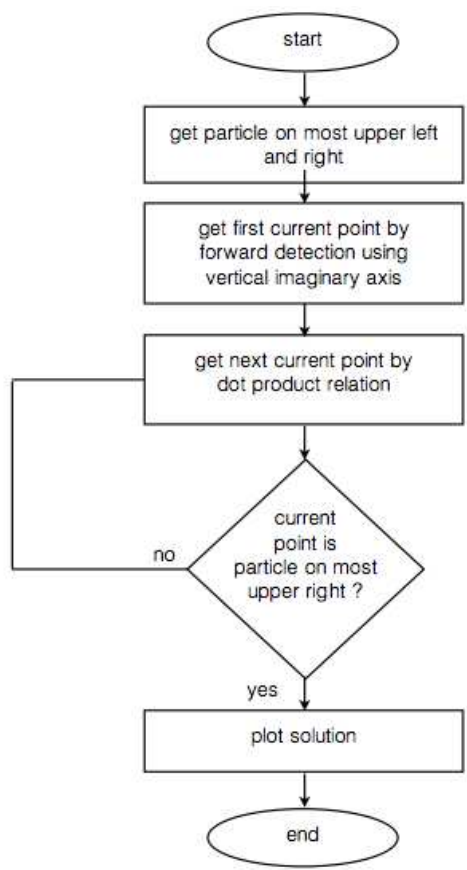}
	\caption{Surface detection algorithm}
	\label{fig:surflow}
	\end{figure}

Once the next surface particle is determined, this particle becomes the current point and the dot product algorithm is used to get the next surface particle. This mechanism is repeated until the upper right particle is detected. The flow chat of the algorithm is shown in Fig. \ref{fig:surflow}.

In some simulations, particles jump off. The algorithm should consider a constraint that would exclude this kind of particles because these particles are actually not part of the wave. The value of particle number density is used as a constraint. Namely If the particle number density is less than $0.4 n_0$, the algorithm will exclude these particles from the surface detection process. 

\section{Test Cases}
	\subsection{Dam break problem}
Two-dimensional dam break problem were simulated. This kind of simulation has been the most common test case in fluid dynamics.

	\subsubsection {Conditions of computation}
The computation domain was set to be $((x,y)| 0 \leq x \leq l, 0 \leq y \leq l)$, where $l$ was taken to be $2$ m. The initial water column was set to be $((x,y)| 0 \leq x \leq 0.5, 0 \leq y \leq 1)$. Rigid boundary condition was applied. The initial velocity and pressure was set to zero for each particle. The distance between particles was set to be $0.02$ m and $2.171$ particles were used in the computation while the simulation is taken until 15 seconds. The initial configuration of the particles is illustrated in Fig. \ref{fig:daminit}.
	\begin{figure}[hbt]
	\centering
	\includegraphics[scale = 0.5]{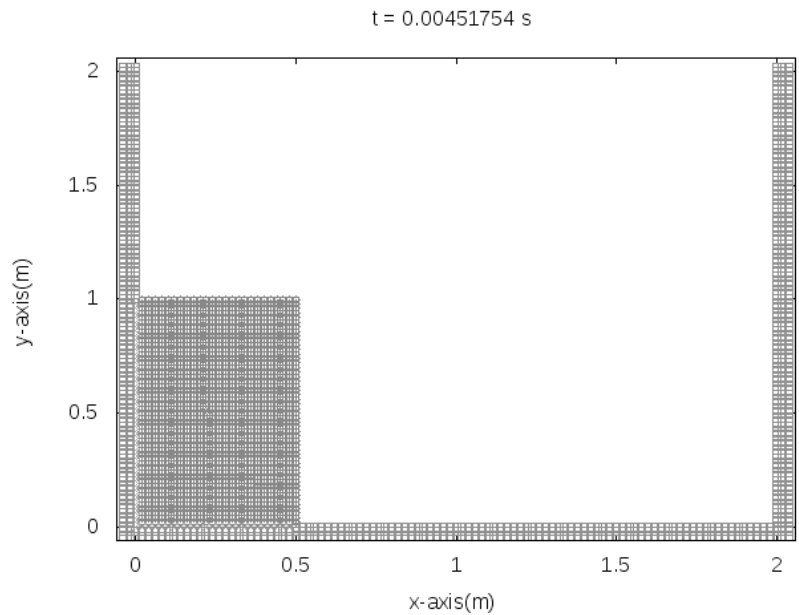}
	\caption{Initial condition of dam break problem}
	\label{fig:daminit}
	\end{figure}

	\subsubsection {Results}
In this implementation, the basic MPS and stochastic MPS were simulated. Stochastic MPS used several $\alpha$ values (see  (\ref{eq:mag})). Namely, $\alpha = 0.003$, $\alpha = 0.001$ and $\alpha = 0.0009$ were simulated. 

The results of the simulation are shown in Fig. \ref{fig:damresult}. Since the time step on each simulation is different, then the comparison was taken on the slightly different time and the main focus of the comparison is the time when wave disappeared . As shown on Fig. \ref{fig:damresult}, at time $12.76s$ It is shown that the wave from the basic MPS already disappeared while the stochastic MPS still survived until several cycle. Thus it can be concluded that the wave which performed by stochastic improvement stronger and survived longer than the basic MPS.

	\begin{figure}[h!]
	\begin{center}$
	\begin{array}{cc}
	\includegraphics[scale = 0.32]{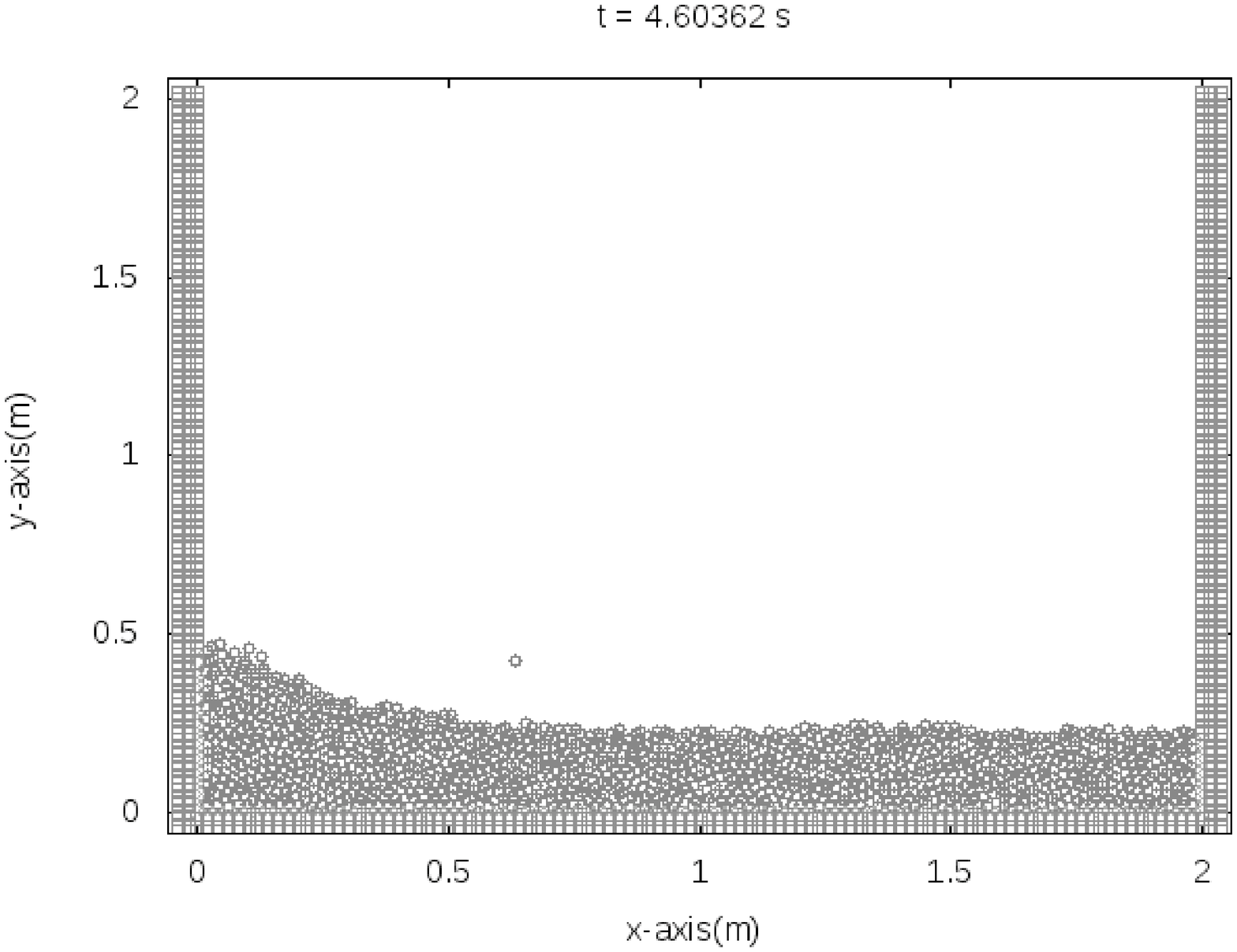} &
	\includegraphics[scale = 0.32]{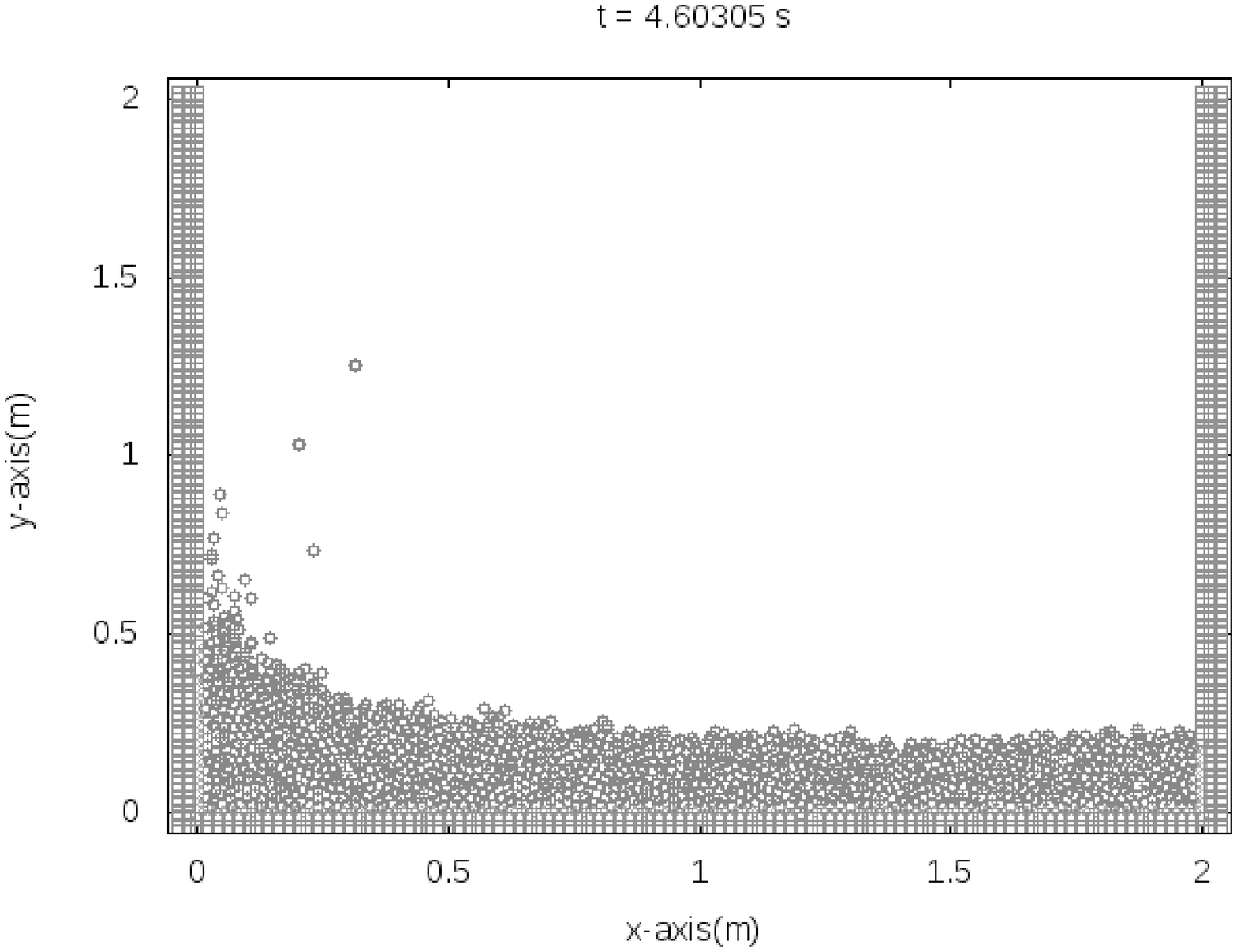}\\

	\includegraphics[scale = 0.32]{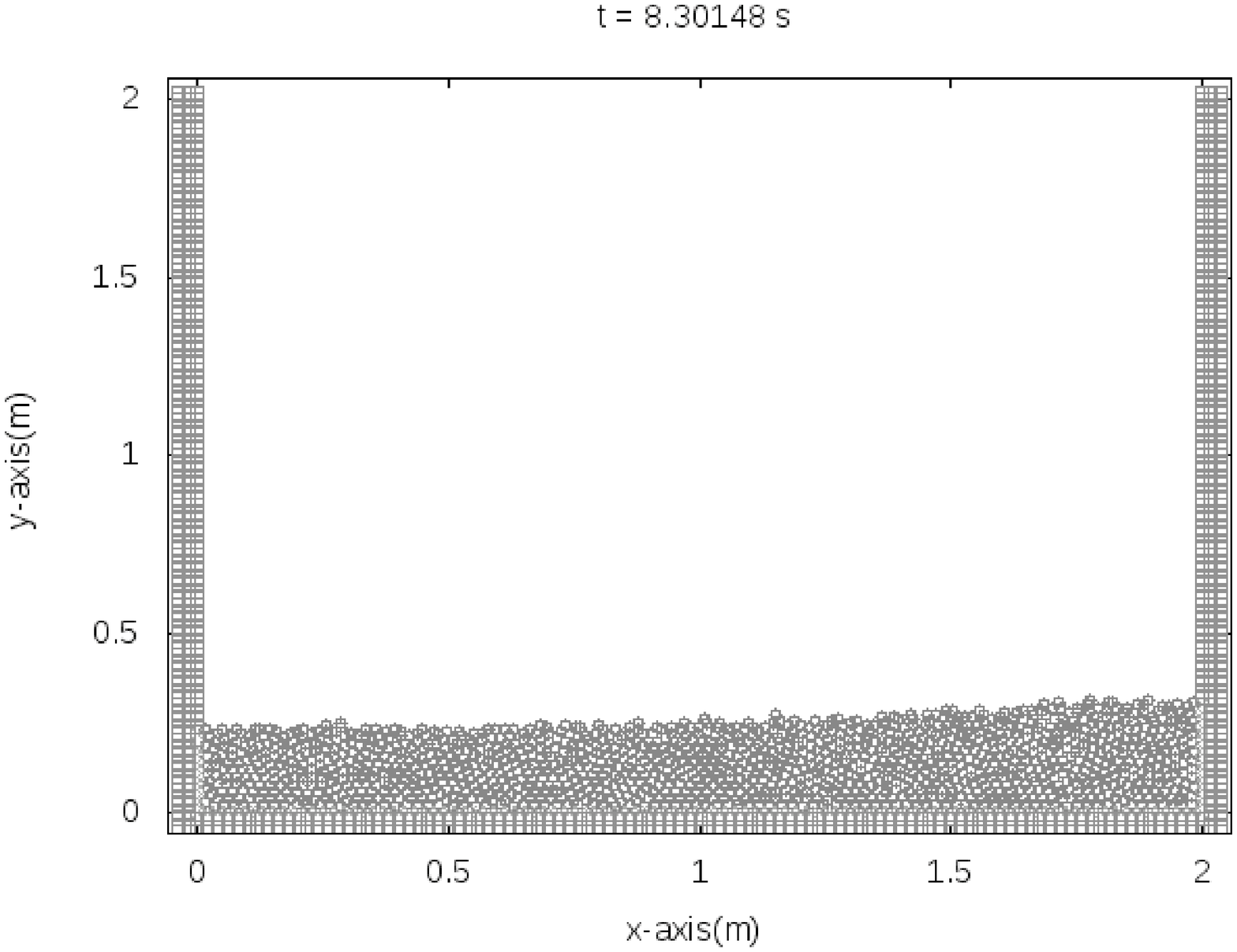} &
	\includegraphics[scale = 0.32]{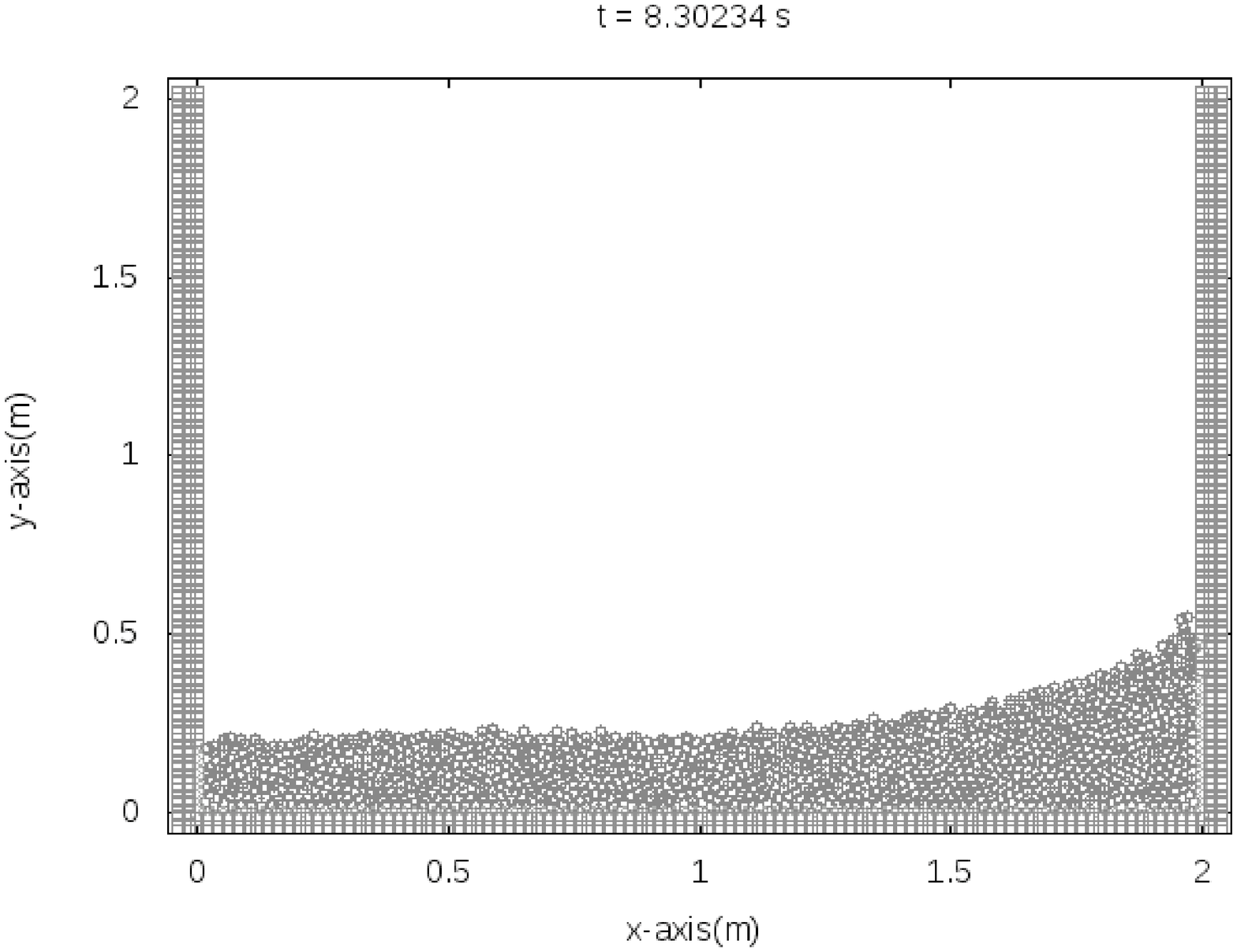}\\

	\includegraphics[scale = 0.32]{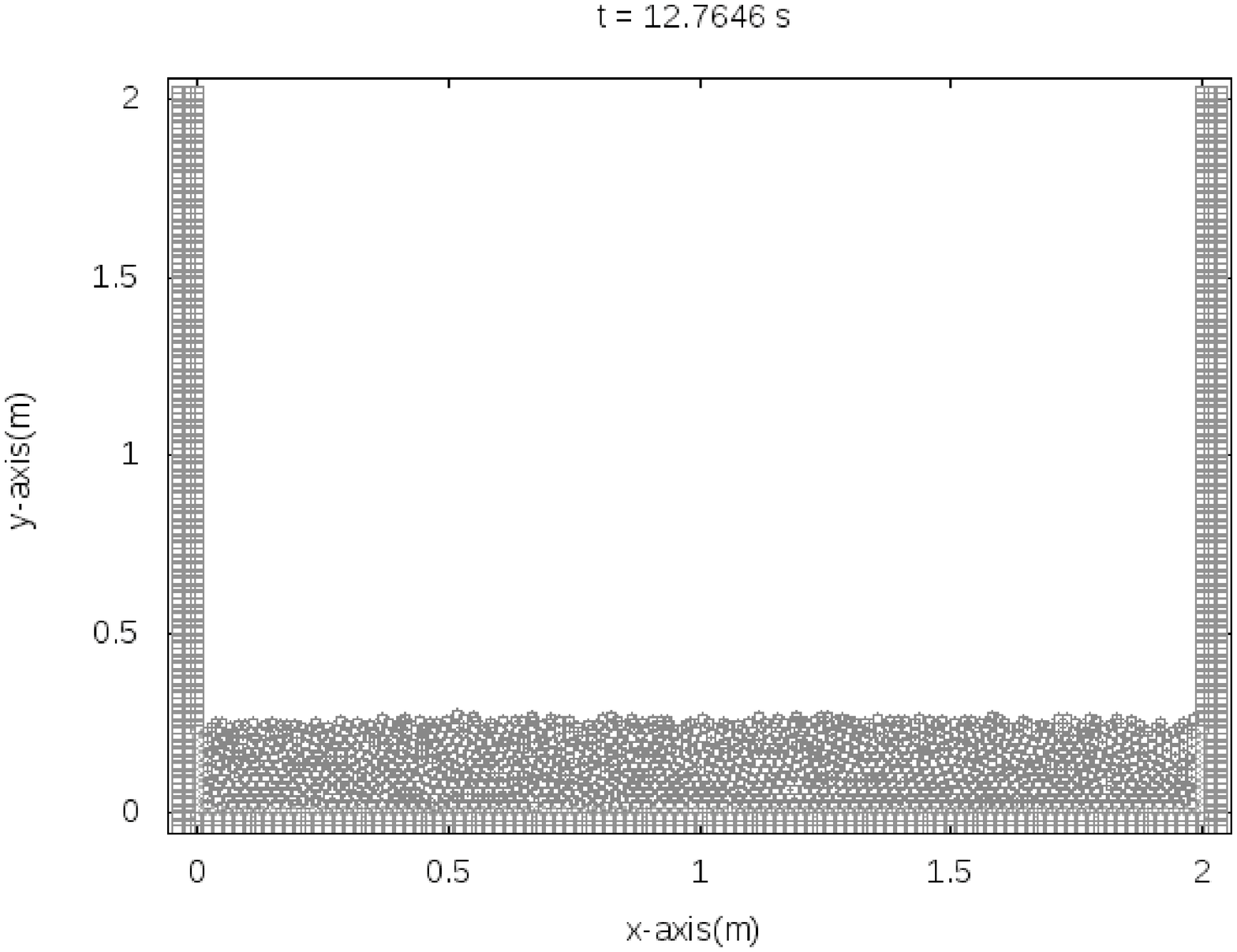} &
	\includegraphics[scale = 0.32]{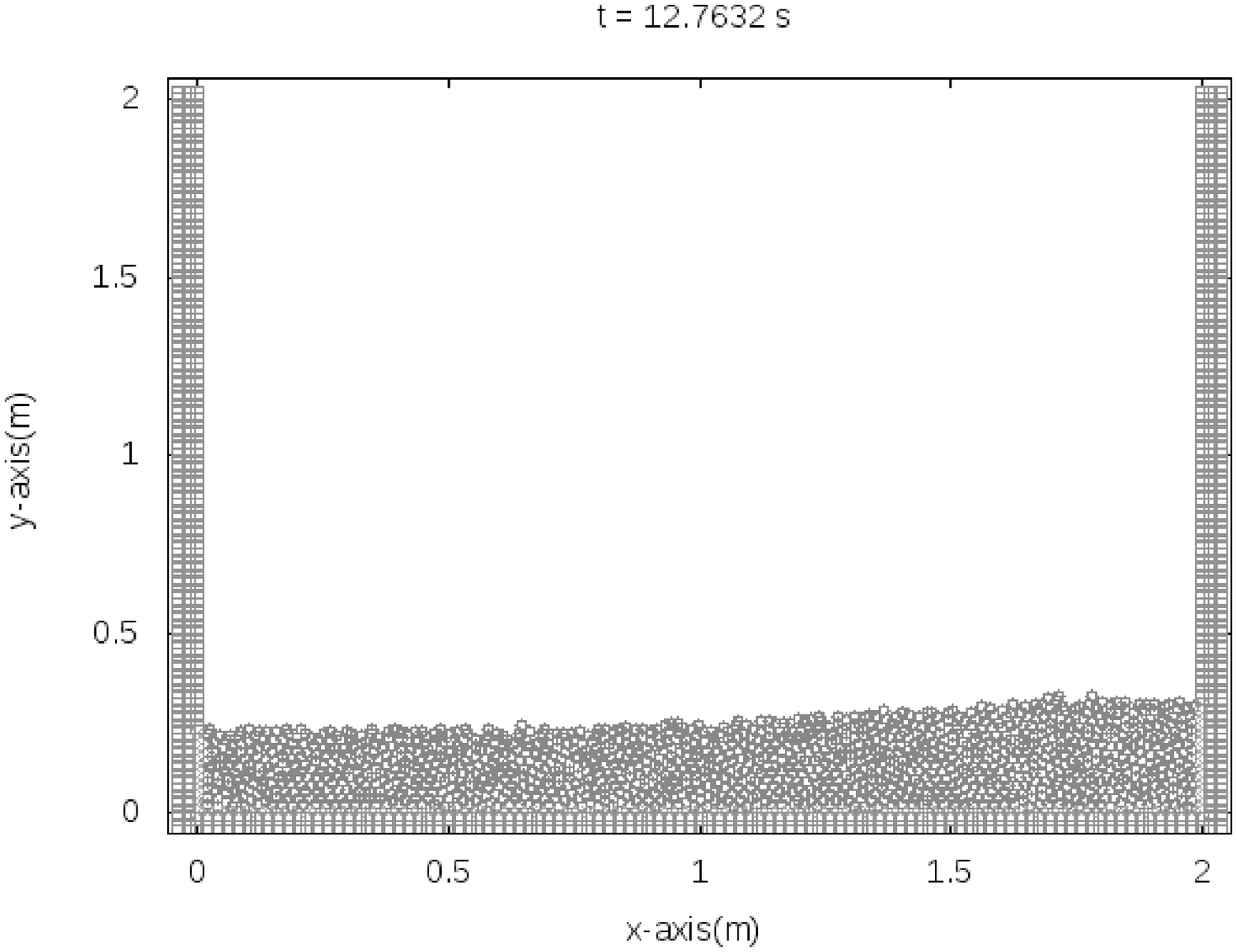}\\

	\end{array}$
	\end{center}
	\caption{Comparison of dam break results between the stochastic MPS with $\alpha = 0.003$ (left) and the basic MPS (right)}
	\label{fig:damresult}
	\end{figure}

The total energy density with respect to time is plotted in Fig. \ref{fig:damenergy}. It is shows that the energy performed by basic MPS has lower state than the stochastic. The oscilation of stochastic energy density is caused by the current algorithm that perform the addition of stochastic velocity only if the total energy on the particular step has lower value than the step before. Since the whole energy system will decrease, the oscilation will occur because of the current stochastic velocity algorithm.

	\begin{figure}[h!]
	\centering
	\includegraphics[scale = 0.4, angle = 270]{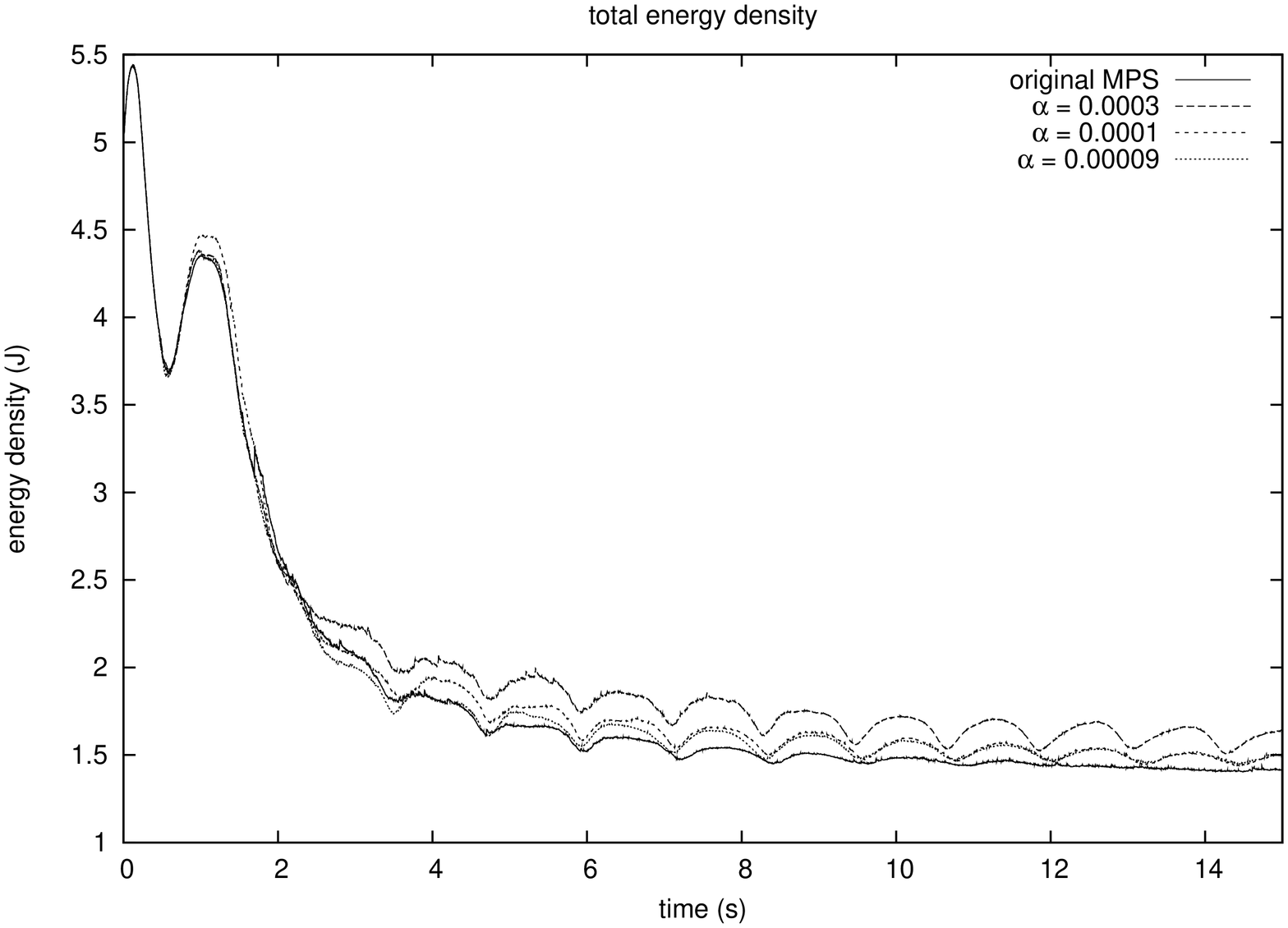}
	\caption{Comparison between basic MPS and the stochastic MPS, $\alpha = 0.003$, $\alpha = 0.001$ and $\alpha = 0.0009$ according to total energy density}
	\label{fig:damenergy}
	\end{figure}

	\subsection{Standing wave in a rectangular tank}
Two-dimensional standing waves in a rectangular tank were simulated. This kind of simulation has been studied by Suzuki \cite{[suzuki]} using Hamiltonian moving particle semi-implicit (HMPS). The results are compared with the analytical solution according to Wu and Taylor \cite{[wu]} based on the water elevation at the center of the tank. In this paper, different approach using surface detection and least square curve fitting is used to determine the amplitude of the wave.
	
	\subsubsection {Conditions of computation}
The computation domain was set to be $((x,y)| 0 \leq x \leq l, 0 \leq y \leq l)$, where $l$ was taken to be equal to the wavelength. Here the wavelength $\lambda$ was $1$ m. The depth of water $h$ was $\lambda/3$. The initial configuration of the free surface is given by
	\begin{equation} 
	y_0 (x) = A \cos[k(x+l/2)].
	\label{eq:etanol}
	\end{equation}
Here $y_0$ is the initial surface displacement, $A = 0.07 \lambda$ is the amplitude and $k$ is the wave number which defined as $2\pi/\lambda$. Periodic boundary conditions were applied, while at the bottom mirror boundary condition was used. The initial velocity and pressure was set to zero for each particle. The distance between particles was set to be $0.02$ m and 834 particles were used in the computation while the simulation is taken until $6$ seconds. The initial configuration of the particles is illustrated in Fig. \ref{fig:initstand}.
	\begin{figure}[hbt]
	\centering
	\includegraphics[scale = 0.45]{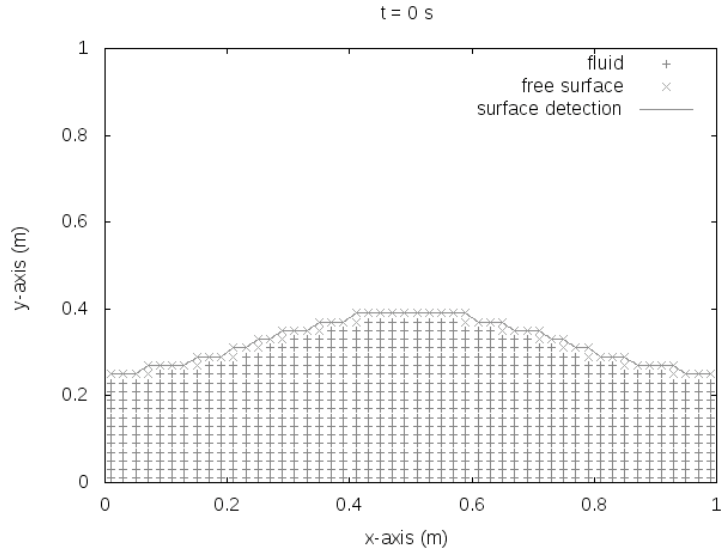}
	\caption{Initial condition of standing wave in a rectangular tank}
	\label{fig:initstand}
	\end{figure}

Surface detection was applied to the solution and the obtained surface was fitted using least square curve fitting. {\emph{GNUPLOT}} curve fitting is used. In this case, since the surface seems closed to quadratic function then the data fitted using simple quadratic function
	\begin{equation}
	y(x) = ax^2 + bx + c,
	\label{eq:fitquad}
	\end{equation} 
where constant $a$, $b$ and $c$ are to determined.

From the result of the least square curve fitting, the amplitude of the wave is calculated on each time step. From the quadratic function (Eq. \ref{eq:fitquad}), the maximum point ($x_{\text{max}},y_{\text{max}}$) is calculated by setting $\frac{dy}{dx} = 0$ which results in
	\begin{equation}
	\begin{cases}
	x_{\text{max}} = \frac{-b}{2a} \\
	y_{\text{max}} = ax_{\text{max}}^2 + b x_{\text{max}} + c.
	\end{cases}
	\label{eq:waveamp}
	\end{equation} 
To estimated the quality of the fitting, average relative error $\bar{\epsilon}$ is used 
	\begin{equation}
	\bar{\epsilon} = \frac{1}{N}\sum_{i = 0}^{N} \left(\frac{|y_{\text{eq}} - y_{\text{est}}|}{y_{\text{eq}}}\right) \times 100 \%.
	\label{eq:rel}
	\end{equation} 
Here, $N$ denotes the number of data, $y_{\text{eq}}$ denotes the fitted value according to  (\ref{eq:fitquad}), while $y_{\text{est}}$ is the $y$ value of the surface detection result. Figure \ref{fig:fitwave} shows the result of surface detection and error calculation.

	\begin{figure}[hbt]
	\centering
	\includegraphics[scale = 0.45]{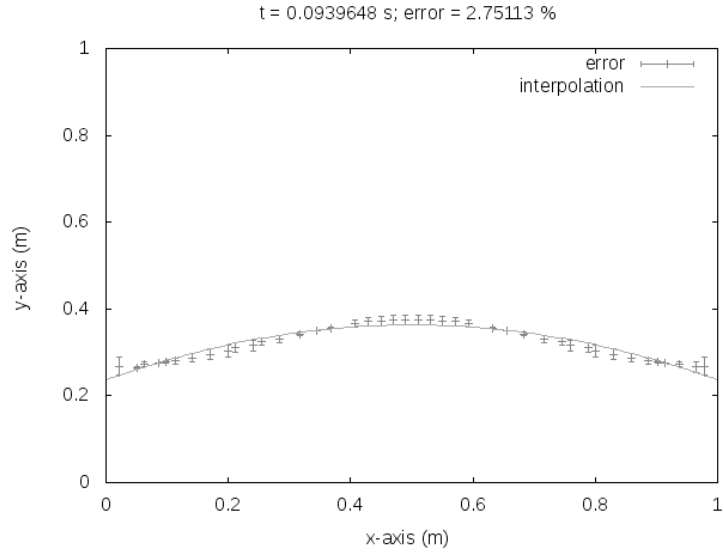}
	\caption{Example of surface detection and curve fitting result}
	\label{fig:fitwave}
	\end{figure}
	
\subsubsection {Results}
In this implementation, the basic MPS and stochastic MPS were simulated. Stochastic MPS used several $\alpha$ values (see  (\ref{eq:mag})). Namely, $\alpha = 0.02$, $\alpha = 0.03$ and $\alpha = 0.04$ were simulated. 

According to the relative error which was less than $4.5\%$, it can be concluded that the curve fitting worked well.
	\begin{figure}[hbt]
	\centering
	\includegraphics[scale = 0.45, angle = 270]{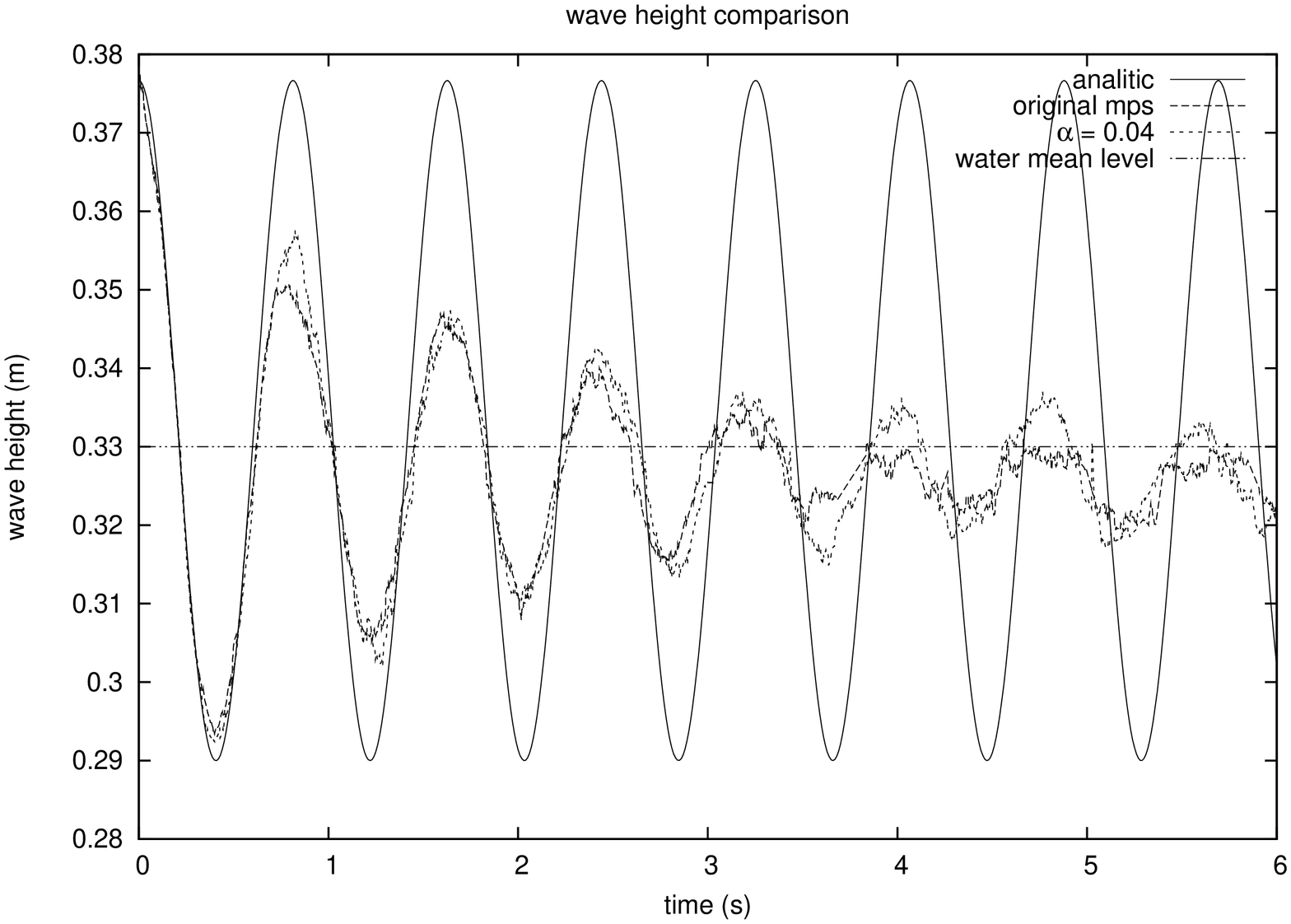} 
	\caption{Comparison between basic MPS, stochastic MPS with $\alpha = 0.04$ and the analytic solution according to wave height (water elevation) in the center of the tank.}
	\label{fig:standerr}
	\end{figure}

The wave height evolution with respect to time is plotted in Fig. \ref{fig:standerr}. It shows that the waves obtained from stochastic MPS survived longer than the basic MPS. The wave from basic MPS decayed rapidly in time, and disappeared already from $t = 3.5$ seconds. In comparison with the analytical solution \cite{[wu]}, the wave height from stochastic MPS was higher and nearer to the analytical solution. 

Ideally in qualitative analysis, if the system has perfect mechanical energy conservation, the amplitude of the wave will not decrease. But since the MPS method suffers from the numerical dissipation, the wave will rapidly decay. Using the stochastic velocity, the rapid decay of the wave could be prevented.
  
	\begin{figure}[!hbt]
	\centering
	\includegraphics[scale = 0.23]{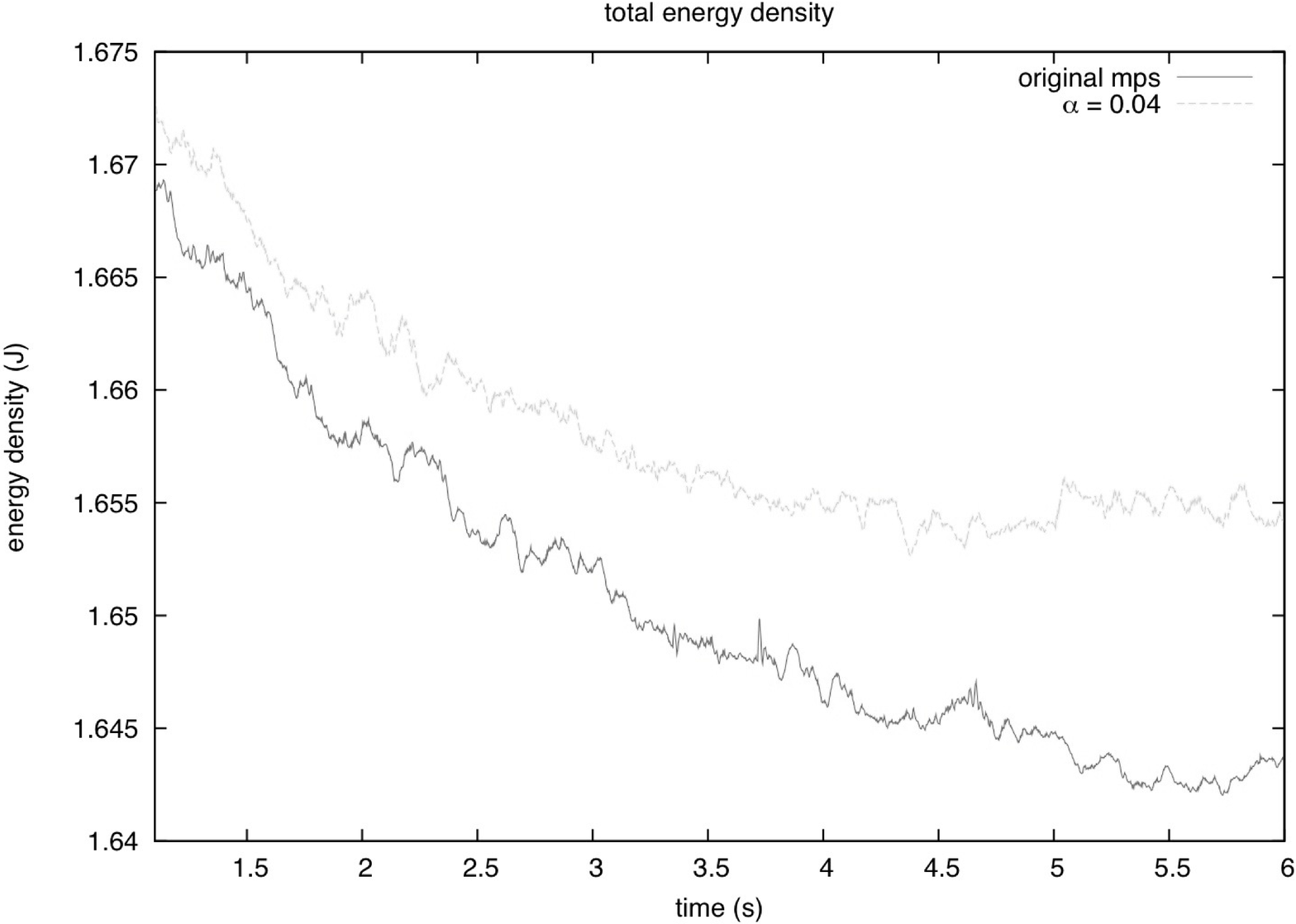} 
		\caption{Comparison between basic MPS and the stochastic MPS using $\alpha = 0.04$ according to total energy density}
	\label{fig:energy}
	\end{figure}

The energy density with respect to time is plotted in Fig. \ref{fig:energy}. It is shown that energy of the basic MPS decayed faster in time compared with the stochastic result (with $\alpha = 0.04$).

However, the stochastic MPS has a drawback that the particles moved to one direction after the wave disappeared. This is caused by the random movement of the particles plus extra velocity added by the stochastic term. The development of stochastic MPS without this drawback is left for further study.
	
\section{Conclusion}
This paper introduces an improvement of moving particle semi-implicit method through the addition of stochastic velocity. Stochastic velocity is added after the explicit step of basic MPS method. The cases that studied were the dam break problem and standing wave in a rectangular tank.  Surface detection and least square curve fitting were used to analyze the waves on the standing wave case. 

On the dam break problem, it was showed that the wave that performed by the stochastic survived longer than the basic MPS. However, the energy density from the stochastic MPS suffers some oscilation due to the stochastic algorithm.

On standing wave case, it was showed that the wave with the stochastic MPS survived longer than the basic MPS. In comparison with the analytical solution \cite{[wu]}, the wave height from stochastic MPS was higher and nearer to the analytical solution even it still far from the ideal system in which the mechanical energy is fully conserved. 

However, on the standing wave case the stochastic MPS has a drawback that the particles moved to one direction after the wave disappeared. This is caused by the random movement of the particles plus extra velocity added by the stochastic term. The development of stochastic MPS without this drawback is left for further study.


\begin{thebibliography}
\small{
\bibitem{[kos1]} Koshizuka S., Oka Y.: {\emph{Moving particle semi implicit method for fragmentation of incompressible fluid}}. Nuclear Science and Engineering, 123(3):421-434, 1996.
\bibitem{[kos3]} Koshizuka S., Nobe A. and Oka Y. : {\emph{Numerical analysis of breaking waves using the moving particle semi-implicit method}}. International Journal of Numerical Methods in Fluids, 26: 751-769, 1998.
\bibitem{[nomura]} Nomura et. al: {\emph{Numerical Analysis of Droplet Breakup Behaviour using Particle Method}}. Journal of Nuclear Science and Technology, 38 (12):1057-1064, 2001.
\bibitem{[khayyer]} Khayyer A., Gotoh H. : {\emph{Modified Moving Particle Semi-implicit method for the prediction of 2D wave impact pressure}}. Coastal Engineering, 56: 419-440, 2009.
\bibitem{[suzuki]} Suzuki Y., Koshizuka S., Oka Y. : {\emph{Hamiltonian moving-particle semi implicit (HMPS) method for incompressible fluid flows}}. Computer methods in applied mechanics and engineering, 196: 2876-2894, 2007.
\bibitem{[szhang]} Zhang et.all, {\emph{An improved MPS method for numerical simulations of convective heat transfer problem}}. International Journal for Numerical Method in Fluids, 51:31-47, 2005.
\bibitem{[wu]} Wu G.X., Taylor R. E. : {\emph{Finite element analysis of two-dimensional non-linear transient water waves}}. Appl. Ocean Res., 16:363-372, 1994.
}

\end{thebibliography}
\end{document}